\documentclass[10pt]{iopart}
\usepackage{amsfonts,amssymb,mathrsfs}

\newcommand{\be}{\begin{equation}}
\newcommand{\ee}{\end{equation}}
\newcommand{\bea}{\begin{eqnarray}}
\newcommand{\eea}{\end{eqnarray}}

\begin{document}

\title{Scalar field mass in generalized gravity}
\author{Valerio Faraoni}
\address{Physics  Department, Bishop's University, 2600 College St., Sherbrooke, 
Qu\'{e}bec, Canada J1M~1Z7}
\eads{\mailto{vfaraoni@ubishops.ca}}
\date{\today} 

\begin{abstract} 
The notions of mass and range of a Brans-Dicke-like scalar 
field in scalar-tensor and $f(R)$  gravity are subject to an 
ambiguity that 
hides a potential trap. We spell out this ambiguity and 
identify a physically meaningful and practical  
definition for these quantities. This is relevant when giving  
a mass to this scalar in order to circumvent experimental limits 
on the PPN parameters coming from Solar System experiments.
\end{abstract} 

\section{Introduction}

The standard theory of gravity, general relativity, is now 
tested rather accurately in the weak-field, slow-motion regime in 
the Solar System \cite{Willbook}, and no deviation from 
Einstein's theory has ever been convincingly demonstrated. On 
the other hand, virtually 
all high energy theories attempting to quantize gravity or 
unifying it with the other interactions predict deviations from 
general relativity. Sometimes, as in the case of the 
simplest string theories, these deviations have to be suppressed 
in order to ensure compatibility with the available experiments 
\cite{bosonic}. At galactic scales, not only general relativity, but even 
Newtonian gravity is doubted in attempts to explain away dark 
matter with MOND and TeVeS theories \cite{MOND, TeVeS}. In a 
modern perspective, looking at alternatives to Einstein's theory 
seems well justified.

The prototypical theory of gravity alternative to general 
relativity is Brans and Dicke's 1961 theory \cite{BD}, which has 
since been generalized to a wider class of scalar-tensor 
theories 
\cite{ST}. The original motivation,  {\em i.e.}, the 
implementation of Mach's principle in 
gravity, is now almost forgotten but these theories have 
nevertheless become 
popular because they incorporate  
key ingredients of string theories, such as the unavoidable 
presence of a dilaton-like gravitational scalar field, and its 
non-minimal coupling to the curvature.

More recently, the discovery that the expansion of the universe 
is accelerated, obtained with the study of type Ia supernovae 
\cite{SN}, has prompted cosmologists to postulate the existence 
of a mysterious form of dark energy with exotic properties to 
explain this acceleration within general relativity  
\cite{Linderresletter}. However, this 
assumption seems rather {\em ad hoc}, and many authors have 
tried to address the issue by assuming instead that the 
Einstein-Hilbert Lagrangian density $R $ receives 
infrared  corrections, being changed to a non-linear function 
$f(R)$ with the non-linearities kicking in at low curvatures 
({\em i.e.}, late in the matter-dominated era) \cite{CCT, CDTT} 
(see \cite{Sergei} for the first models compatible with the 
post-Newtonian constraints in the Solar System). In this 
framework,  the 
cosmic acceleration observed would be the first signal of 
deviations from Einstein's theory. This ``modified'' or 
``$f(R)$''  
gravity contains a massive scalar degree of freedom in 
addition to the familiar massless graviton. $f(R)$ gravity 
turns out to be  equivalent to a Brans-Dicke theory 
\cite{STequivalence}.

Usually, the Brans-Dicke-like scalar field of scalar-tensor 
theories is given a mass in order to make it short-ranged and 
evade the weak-field constraints coming from Solar System 
experiments. The concepts of mass and 
range used are subject to an ambiguity that hides a potential 
trap 
and should be clarified. The purpose of this paper is to spell 
out this ambiguity and clarify the use of these quantities, 
especially in relation with the Parametrized Post-Newtonian 
formalism used to constrain scalar-tensor gravity 
\cite{Willbook}. We are not 
aware of such a discussion in the vast literature on 
scalar-tensor gravity.

Here we use Brans-Dicke theory \cite{BD} for the sake of 
illustration, but the discussion  can be extended to more 
general scalar-tensor theories. Before proceeding, it is useful  
to recall the basic ingredients of Brans-Dicke gravity 
that we use in this paper. The 
gravitational field is described by both a metric tensor 
$g_{ab}$ and a scalar field $\phi$ appearing in the (Jordan frame) 
action~\footnote{The original Brans-Dicke theory \cite{BD} did 
not include a potential $V(\phi)$, and many authors prefer to 
reserve the name ``Brans-Dicke theory'' to the case in which 
$V=0$. Then, the theory described by~(\ref{Jaction}) could be 
called 
``scalar-tensor theory with constant $\omega$ and linear 
coupling function''. For economy of terminology,  we still refer 
to~(\ref{Jaction}) as a Brans-Dicke action.}
\begin{equation} 
S_{BD} = \frac{1}{16\pi}\int d^4x \, \sqrt{-g} \left[ \phi R 
-\frac{\omega}{\phi}\, g^{ab}\nabla_a \phi\nabla_b \phi -V(\phi) 
 +{\cal L}^{(m)} \right] \;, \label{Jaction}
\end{equation}
where $\omega $ is the constant Brans-Dicke parameter, $V(\phi)$ 
is a scalar field potential, and ${\cal L}^{(m)}$ is the matter 
Lagrangian density. We adopt the notations of 
Ref.~\cite{Wald}, but we retain Newton's constant $G$ in our  
equations without setting it to unity. Note that the 
Brans-Dicke field $\phi$ 
is identified (apart, possibly, from a numerical coefficient)  
with the inverse effective gravitational coupling 
$G^{-1}=m_{Pl}^2$ (where $m_{Pl}$ is the Planck mass) and has 
therefore the dimensions of  a mass squared. We 
restrict ourselves to positive values of $\phi$  to ensure that the graviton carries 
positive kinetic 
energy, and to the range of values of the Brans-Dicke 
parameter $2\omega+3>0$ (the value $\omega=-3/2$ 
corresponds to a barrier that can not be crossed by a continuous 
change of this parameter, a theory  with a non-dynamical 
scalar field $\phi $, a Cauchy problem that is ill-posed with most forms of matter   
\cite{TremblayFaraoni, review}, and an unphysical 
dependence of the metric on derivatives of the matter fields 
of order higher than second \cite{BarausseSotiriouMiller}).   
The Brans-Dicke field equations are \cite{BD}
\begin{eqnarray} 
&& G_{ab}=\frac{8\pi }{\phi}\, 
T^{(m)}_{ab}+\frac{\omega}{\phi^2} 
\left( \nabla_a \phi \nabla_b \phi -\frac{1}{2}\, g_{ab} 
\nabla^c \phi\nabla_c \phi \right) \nonumber\\
&&\nonumber \\
&& + \frac{1}{\phi}\left( \nabla_a \nabla_b \phi -g_{ab}\Box 
\phi \right) 
-\frac{V}{2\phi}\, g_{ab} \;, \label{BDfieldeq1}\\
&&\nonumber \\
&& \Box \phi=\frac{1}{2\omega +3} \left( 8\pi T^{(m)} +\phi\, 
\frac{dV}{d\phi}-2V \right) \;, \label{BDfieldeq2} 
\end{eqnarray}
where $T_{ab}^{(m)}=\frac{-2}{\sqrt{-g}} \, \frac{\delta }{\delta 
g^{ab}} \left( \sqrt{-g} \, {\cal L}^{(m)} \right)$ is the 
matter  energy-momentum tensor and $T^{(m)}$ is its trace.

The Jordan frame action~(\ref{Jaction}) can be mapped into its 
Einstein frame representation by means of the conformal 
transformation
\be \label{conformalmetric}
g_{ab}\rightarrow \tilde{g}_{ab}=\Omega^2 \, g_{ab} \;, 
\;\;\;\;\;\;\; \Omega=\sqrt{G\phi} \;, 
\ee
accompanied by the scalar field redefinition
\be\label{conformalscalar}
\phi \rightarrow \tilde{\phi}=\sqrt{ \frac{2\omega +3}{16\pi G}} 
\, \ln \left( \frac{\phi}{\phi_*} \right) 
\ee
to make the kinetic energy of the new scalar assume the 
canonical form, and where $\phi_* $ is an irrelevant constant. 
Note that the dimensions of the new scalar field $\tilde{\phi}$ 
are those of a mass. The Einstein frame form of the action is
\begin{equation} 
S_{BD}= \int d^4x \, \sqrt{-\tilde{g}} \left[ \frac{ 
\tilde{R}}{16\pi G}  
-\frac{1}{2}\, \tilde{g}^{ab}\tilde{\nabla}_a 
\tilde{\phi} \tilde{\nabla}_b \tilde{\phi} -U\left( 
\tilde{\phi} \right)  +\frac{{\cal L}^{m}}{\left( G\phi\right)^2}  
\right]   \;,\label{Eaction} 
\end{equation}
where the new scalar field potential is 
\be
U\left( \tilde{\phi} \right)=\frac{ V\left[ \phi 
\left( \tilde{\phi} \right) \right] }{ \left( G\phi 
\left(\tilde{\phi} \right) \right)^2 } =
V \left[ \phi \left( \tilde{\phi} \right) \right] \exp \left( 
-8\sqrt{ \frac{ \pi G}{2\omega+3} } \,  \tilde{\phi} \right) 
\ee
and a tilde denotes geometrical quantities associated with  
$\tilde{g}_{ab}$.

\section{What is the mass of a Brans-Dicke scalar field?}

Let us begin with the Jordan frame description of the theory.   
According to current 
terminology used for all kinds of scalar 
fields, in the presence of a 
non-trivial potential $V$ the Brans-Dicke scalar $\phi$ 
is massive, with effective mass $\mu$ given by
\be \label{mu}
\mu^2 (\phi) \equiv \frac{d^2V}{d\phi^2} \;.
\ee
However, it is already clear that this definition is 
problematic because $\mu$ is dimensionless ($V$ has the 
dimensions of an energy density, mass$^4$, and so does 
$\phi^2$). Apart from the dimensions, in Brans-Dicke theory 
other  definitions of effective mass are clearly possible. The 
Einstein frame effective mass 
$\tilde{m}$ is given by
\be \label{mtilde}
\tilde{m}^2 (\tilde{\phi} ) \equiv 
\frac{d^2U}{d\tilde{\phi}^2} \;.
\ee
When quantizing linearized scalar-tensor gravity in a canonical approach, the 
Einstein frame variables are necessary: it is the Einstein frame and not the Jordan 
frame  metric perturbation  that must be identified with the physical 
graviton and 
leads to the correct propagator. The 
propagator for the Einstein frame scalar field $\tilde{\phi}$ 
 in the absence of matter  yields again $\tilde{m}$ as the 
scalar field mass \cite{Eframequantization}.

We want to express $\tilde{m}$ in terms of $\phi$ instead 
of $\tilde{\phi}$. Eq.~(\ref{mtilde}) yields 
\begin{eqnarray}
&& \tilde{m}^2 = \frac{d^2}{d\tilde{\phi}^2} \left[ 
\frac{V(\phi( \tilde{\phi}) ) }{\left( G \phi( \tilde{\phi}) 
\right)^2}  \right]   =
\frac{1}{G^2 \phi^2}\left[ \left( 
\frac{dV}{d\phi}-\frac{2V}{\phi} \right) 
\frac{d^2\phi}{d\tilde{\phi}^2} \right. \nonumber \\
&& \nonumber \\
&& \left. +\left( 
\frac{6V}{\phi^2}-\frac{4}{\phi}\, \frac{dV}{d\phi} 
+\frac{d^2V}{d\phi^2} \right)\left( 
\frac{d\phi}{d\tilde{\phi}} \right)^2 \right] 
\end{eqnarray}
and, using
\be
\frac{ d\phi}{d\tilde{\phi}} =\left( \frac{16\pi 
G}{2\omega+3} \right)^{1/2} 
\, \phi \;, \;\;\;\;\;\;\;\;
\frac{ d^2\phi}{d \tilde{\phi}^2} =\frac{16\pi 
G}{2\omega+3} 
\, \phi \;, 
\ee
it leads to
\be \label{mtilde2}
\tilde{m}^2(\phi)=\frac{16\pi }{\left( 2\omega+3 
\right) G\phi} \left( \frac{4V}{\phi}-3\, 
\frac{dV}{d\phi}+\phi\, \frac{d^2V}{d\phi^2} \right) \;.
\ee
Let us now see, in the Jordan frame,  a third possible 
definition of effective mass of $\phi$. In general, for a 
Klein-Gordon field $\varphi$ satisfying  
$\Box \varphi-\frac{dW}{d\varphi}=S$, where $S$ 
is a source term, the effective mass is defined by 
$m^2_{\varphi}(\varphi) = d^2W/d\varphi^2$. In the case of 
a quadratic potential $W(\varphi)=\frac{\alpha}{2} \, 
\varphi^2 + \beta \varphi + \gamma $ (with $\alpha, \beta$, and 
$\gamma$ constants), this leads to a  constant  
$m_{\varphi}$, and the definition certainly makes sense around a minimum of the 
potential~\footnote{Already the presence of a tadpole term 
corresponding to $\beta 
\neq 0$ leads  to non-equilibrium and then the quantity $m_{\varphi}^2$ 
can not be 
interpreted as a true mass squared.},  {\em i.e.}, for $\beta=0$ and 
$\alpha>0$. It 
is the common use of  this definition that  inspires eq.~(\ref{mu}).  However, the 
Brans-Dicke field $\phi$ does not satisfy the usual  
Klein-Gordon equation, but obeys eq.~(\ref{BDfieldeq2}) 
instead, in which the potential $V$ appears in an  
unconventional way. Eq.~(\ref{BDfieldeq2}) can be turned into 
the 
usual  Klein-Gordon form by introducing an effective potential 
$V_{eff}(\phi)$ satisfying
\begin{eqnarray} \label{KleinGordonform}
&& \Box \phi -\frac{dV_{eff}}{d\phi} =\frac{8\pi 
T^{(m)}}{2\omega +3} \;, \\
&& \nonumber\\
&& \frac{dV_{eff}}{d\phi}=\frac{1}{2\omega+3}\left( 
\phi \, \frac{dV}{d\phi} -2V \right) \;.
\end{eqnarray}
Integration with respect to $\phi$ then yields 
\be\label{Veff}
V_{eff}(\phi)=\frac{1}{2\omega +3} \left[ \phi V(\phi) -3\int 
d\phi \, V(\phi) \right] \;.
\ee
Note that the addition of a constant to $V_{eff}(\phi)$ 
has no effect on the dynamics of $\phi$, while adding 
a constant to $V(\phi)$ does (more on this later). While 
$V$ has the dimensions of an energy density, $V_{eff}$ 
has the dimensions of an energy density times  a mass 
squared. 

The relation~(\ref{Veff}) between $V$ and $V_{eff}$ is 
rather peculiar: if $V$ is even [odd], $V_{eff}$ is odd 
[even], and vice-versa. If $V$ is a polynomial of 
degree $n$ in $\phi$, then $V_{eff}$ is  a polynomial of 
degree $(n+1)$. If $V$ is positive-definite or bounded, 
the same is not true in general for $V_{eff}$.

Now, with the example of  the Klein-Gordon equation in 
mind, one can define a third effective mass $m$ for the 
field $\phi$ by\footnote{This definition is used  
also in \cite{Helbig}.}
\be\label{m}
m^2(\phi) \equiv \frac{d^2 
V_{eff}}{d\phi^2}=\frac{1}{2\omega+3} \left( \phi \, 
\frac{d^2V}{d\phi^2}-\frac{dV}{d\phi} \right) \;.
\ee
A  sufficient condition for $V_{eff}$ to have a minimum at $\phi_0$ is 
$V_0'=2V_0/\phi_0$ and $ 
V_0''>\frac{V_0'}{\phi_0}=\frac{2V_0}{\phi_0^2}$.
As for the definition~(\ref{mtilde}) of 
$\tilde{m}$, it  is also inspired by  the Klein-Gordon example, but it seems that a 
proper use of the Klein-Gordon analogy supports the definition (\ref{m}) rather 
than~(\ref{mtilde}).

We now have  three different definitions~(\ref{mu}),  (\ref{m}), 
and (\ref{mtilde}) of 
effective mass of  $\phi$, the first two given using  Jordan frame 
quantities and the third given in the Einstein frame. At least formally, we can 
retain these definitions even away from minima of the respective (effective) 
potentials, altough their interpretation as ``masses'' would fail from a strict 
particle physics point of view.

In principle, infinitely many definitions 
are possible, corresponding to the infinitely many 
conformal frames that can be defined based on different choices 
of the conformal factor $\Omega (\phi) $ in the 
conformal transformation of the 
metric $\tilde{g}_{ab}=\Omega^2 \, g_{ab}$ and the 
corresponding 
potential $U(\tilde{\phi})=V(\phi(\tilde{\phi}))/\Omega^4 $. 
However, to be practical, we will focus on the three 
definitions  given above.

Let us now see how bad the ambiguity in the 
concept of ``mass of $\phi$'' is by considering various 
potentials. These are not mere examples: as will be 
clear below, they correspond to the only cases in which 
serious ambiguities arise. 

\subsection{Constant potential $V=V_0$}

In this case it is
\be
\mu=0 \;, \;\;\;\;\; 
m=0 \;, \;\;\;\;\;
\tilde{m}=8\sqrt{ \frac{\pi  V_0}{(2\omega+3)G }} \, 
\phi^{-1} \;.
\ee
The scalar $\phi$ has infinite range according to all 
three definitions if $V_0=0$, but has finite range 
according to~(\ref{mtilde}) and infinite range according 
to both (\ref{mu}) and~(\ref{m}) if $V_0\neq 0$. 

\subsection{Linear potential $V=\alpha \phi + V_0 $}

In this case, it is 
\be
\mu=0 \;, \;\;\;\;\;
m=\sqrt{\frac{-\alpha}{2\omega+3} } \;, \;\;\;\;\;
\tilde{m}=4\sqrt{\frac{\pi  
(\alpha+4V_0/\phi ) }{(2\omega+3) 
G \phi} } \;. 
\ee
A natural choice would seem to be $\alpha >0$ which, for 
the assumed ranges $\phi>0$ and 
$2\omega+3>0$ of the scalar and of the Brans-Dicke 
parameter, leads to a potential $ V$ that is bounded 
from below. However, this leads to a tachyonic scalar 
according to the definition~(\ref{m}), which is based on the 
dynamical equation actually obeyed by $\phi$. Aside from 
this problem, $\phi$ seems to have infinite range 
according to $\mu$ but finite range according to  
$\tilde{m}$.

\subsection{Quadratic potential $V=\frac{\alpha}{2}\,  
\phi^2 + \beta $}

In this case it is
\be
\mu=\sqrt{\alpha} \;, \;\;\;\;\;
m=0 \;, \;\;\;\;\;
\tilde{m}=8\sqrt{ \frac{\beta}{ (2\omega+3)G }}\,  \phi^{-1} 
\;. 
\ee
If $\alpha>0$, then $\phi$ has finite range according 
to $ \mu$ but infinite range according to $m$, and 
finite or infinite range according to the Einstein frame 
$\tilde{m}$ if $\beta>0$ or 
$\beta=0$, respectively. Negative values of $\beta$ are ruled 
out by 
the requirement that the field is not tachyonic in the 
Einstein frame. {\em A priori}, we identify 
here the possibility of a field being tachyonic in one 
conformal frame but not in the other, if the definitions given 
above make sense.

At this point, one may want to find {\em all} the 
instances in which $\tilde{m}=0$ while $m\neq 0$, which 
is equivalent to finding all the functions $V(\phi)$ 
satisfying the ODE
\be\label{ODE}
\frac{d^2V}{d\phi^2}-\frac{3}{\phi}\, 
\frac{dV}{d\phi}+\frac{4V}{\phi^2} =0 
\ee
with $m \neq 0$. Eq.~(\ref{ODE}) has the general solution
\be\label{cccc}
V(\phi)=\phi^2 \left[ A+B\ln \left( \frac{\phi}{\phi_0} 
\right) \right]
\ee
with $A, B$, and $\phi_0>0$ arbitrary constants. For $B=0$ 
one recovers the case of a harmonic potential already 
considered, in which both $m$ and $\tilde{m}$ vanish. 
Therefore, the {\em only} case in which 
$\tilde{m}=0$ while $m\neq 0$ is that of $V(\phi)$ given by 
eq.~(\ref{cccc}) with $B\neq 0$. Moreover, 
the {\em only} case in which both $m$ and $\tilde{m}$ 
vanish is that of  a purely quadratic potential $B=0$ 
(including, of course, the trivial case $A=0$).
The fact that the choice $V=\mu^2\phi^2/2 $ makes the potential 
disappear from the field equation~(\ref{BDfieldeq2}) was 
noticed, but not pursued, in studies of the phase space of 
spatially  homogeneous and isotropic scalar-tensor cosmology 
\cite{SantosGregory, mybook, phasespace}.

\subsection{Potential $V=B   \phi^2 \ln 
\left(\phi/\phi_0 \right)$}

In this case it is
\begin{eqnarray}
\mu & = &  \sqrt{B\left[ 2\ln \left( \frac{\phi}{\phi_0} 
\right)+3 \right]}  \;, \\
&&\nonumber \\
m &= & \sqrt{ \frac{2B \phi}{ 2\omega+3 }}   \;, \\
&&\nonumber\\
\tilde{m} & = & 0 \;. 
\end{eqnarray}
If $B>0$, then $\phi$ has finite range in the Jordan 
frame (according to both $\mu$ and $m$) but infinite 
range in the Einstein frame ($\mu$ could also become imaginary). 

\subsection{Consequences}

Needless to say, the ranges (or masses) of the 
Brans-Dicke scalar can be quite 
different according to which of the three 
definitions~(\ref{mu}), (\ref{m}), or (\ref{mtilde}) is 
used, and care must be taken when using these 
quantities. This is particularly important when 
discussing violations of the experimental constraints in 
the  Parametrized Post-Newtonian (PPN) formalism. For a 
free Brans-Dicke field, the experimental limit on the 
Brans-Dicke parameter is $\omega \geq 40000$ 
\cite{BertottiIessTortora}. To circumvent this 
constraint, a potential is usually given to $\phi$  to 
make it  short-ranged, thereby rendering local (terrestrial 
and Solar System) experiments  insensitive~\footnote{There are no 
experimental  constraints on gravity at 
distances smaller than 0.2~mm.} to $\phi$.  This was done, historically, 
when it was recognized 
that one of the first string theories, bosonic string theory, 
has a low-energy limit corresponding to $\omega=-1$ 
Brans-Dicke gravity \cite{bosonic, TaylorVeneziano}. More 
recently, 
a large amount of literature has been devoted to $f(R)$ 
theories of gravity described by the action
\be\label{f(R)action}
S=\frac{1}{16\pi G} \int d^4 x \, \sqrt{-g} \left[ f(R) 
+ {\cal L}^{(m)} \right] \;,
\ee
in both the metric and Palatini approach (see 
\cite{review, otherreviews} for 
reviews). 
It is well known that these theories are equivalent to  
an $\omega=0$ (for the metric case) or an 
$\omega=-3/2$ (for the Palatini case) Brans-Dicke 
theory \cite{STequivalence}. The constraint on $\omega$ 
is circumvented by giving an effective mass to the 
Brans-Dicke scalar through the chameleon mechanism 
\cite{review, otherreviews}. Further, $\omega=0$ 
Brans-Dicke theory is singled out by the area metric approach 
unifying metric and scalar field into a single geometric 
structure \cite{PunziSchullerWohlfart}. In these, and in 
similar, cases one needs to be clear 
on what is meant by ``mass'' and ``range'' of $\phi$. First, 
let us focus on the Jordan frame description of 
scalar-tensor gravity. The 
range of $\phi$ is determined by the equation of motion 
satisfied 
by $\phi$ (in the Jordan frame, eq.~(\ref{BDfieldeq2})). 
This can be written in the Klein-Gordon 
form~(\ref{KleinGordonform}) and, in the weak-field, 
slow-motion regime appropriate to the low-density Solar 
System  environment, it becomes
\be
\frac{1}{r^2} \, \frac{d}{dr} \left( r^2 
\frac{d\phi}{dr} \right) -\frac{dV_{eff}}{d\phi} \simeq 
0
\ee
with the further assumption of spherical symmetry 
$\phi=\phi(r)$. The 
effective mass is determined by expanding 
$dV_{eff}/d\phi $ around the present value $\phi_0$ of 
$\phi$,
\be\label{Yukawa}
\frac{dV_{eff}}{d\phi} =
\frac{dV_{eff}}{d\phi}\left. \right|_0 + 
\frac{d^2V_{eff}}{d\phi^2} \left. \right|_0 \phi + \, 
...=m_0^2\phi  + \, ...
\ee
if $V_{eff}$ has  a minimum at $\phi_0$, and where $m_0 
\equiv m(\phi_0)$. Then, eq.~(\ref{Yukawa}) has the 
usual Yukawa solution $\phi(r) \propto \mbox{e}^{-m_0 
r} /r$ with range $m_0$ determined {\em by the 
definition~(\ref{m}), and not by the 
definition~(\ref{mu}) of $\mu$ or 
(\ref{mtilde}) of $\tilde{m}$}. This fact singles out the mass 
$m$ and the corresponding range of $ \phi $ to be used in the 
PPN analysis. 
This also leads to some worries  because, following 
the current terminology and usage, violations of the 
PPN limits are suppressed by giving a short range to 
$\phi$ {\em according to the definition~(\ref{mu}), not 
(\ref{m})} ({\em e.g.}, \cite{Fizievetal}).  If this potential is 
quadratic, 
$ V(\phi)=\frac{\alpha}{2}\, \phi^2+\beta$, the range of 
$\phi$ according to the correct definition~(\ref{m}) in 
the Jordan frame, is still infinite! And it will still 
be infinite also in the Einstein frame if $\beta=0$.  We 
conclude that, in the Jordan frame, one must be careful to 
choose $m$ instead of $\mu$ as the $\phi$-mass.

Let us turn now to the Einstein frame picture. The discussion 
above leads to the potential difficulty that, if  $\beta\neq 0$ 
in the potential $ V(\phi)=\frac{\alpha}{2}\, \phi^2+\beta$,  
the fact 
that $\phi$ 
has finite or infinite range and evades or not the PPN 
constraints, seems to depend on the conformal 
representation of the theory which is adopted.  The 
other potentially dangerous situation is that of the 
potential $V(\phi) \propto \phi^2 \ln ( \phi/\phi_0)$, 
which yields a massive field according to $\mu$ and $m$ 
but a free field in the Einstein frame 
$\tilde{m}=0$. 

The violation or non-violation of the PPN constraints seems to 
depend on 
the conformal representation of the theory adopted. 
However, it is still true that a free field in the Jordan 
frame, defined by $V\equiv 0$, remains a free field in the  
Einstein frame ($U \equiv 0$). Moreover, from a pragmatic point 
of view, the  problem is not so serious because 
the calculation of the PPN parameters and metric potentials is 
done in the Jordan frame (see, {\em e.g.}, \cite{Willbook, 
DamourFarese})~\footnote{This fact happens to be consistent  
with the fact that the physical choice of 
mass  that we identify as physical is given by eq.~(\ref{m}) in 
the Jordan frame, but it is just common practice to use the 
Jordan frame to perform the PPN limit and there is no fundamental 
reason to prefer this conformal frame.}. For this 
reason, we will not discuss this 
issue further. We only  remark that these occurrences do not 
point to a failure of the physical equivalence between the two 
frames; they simply mean 
that care must be taken in defining and calculating quantities 
correctly when going to the Einstein frame 
representation. At least at the classical level, the 
two frames are physically equivalent representations of 
the same theory \cite{Dicke, Flanagan, FaraoniNadeauconftransf}. 
It is however, true, that certain features of 
gravitational theories may not be formulated in a 
representation-independent way, a problem that has been 
discussed elsewhere \cite{STV}.\\

When trying to understand the issue of mass, one should keep in 
mind that this concept relies on a scalar field potential and 
its expansion around a minimum. Now, in general, a potential with a given 
functional form $V(\phi)$ in the Jordan frame corresponds 
to a very different functional form $U\left(\tilde{\phi} 
\right)=V(\phi ( \tilde{\phi}))/ (G\phi)^2$ in the Einstein 
frame. 
For example, a purely quadratic  $V=\mu^2\phi^2/2$ corresponds 
to 
the potential $U\left( \tilde{\phi} 
\right)=\frac{\mu^2\phi_0^2}{2G^2} \, \mbox{e}^{8\sqrt{ 
\frac{\pi 
G}{2\omega + 3}}\, \tilde{\phi} } $. It is well known that 
switching to the Einstein frame representation is a 
mathematical technique which allows one to 
find exact solutions of the field equations that would be 
harder, or impossible to find in the Jordan frame. This 
technique  
is based on the knowledge of exact solutions of general 
relativity. However, these solutions correspond to rather odd 
potentials when mapped back into the Jordan frame. The point is 
that the functional form of the scalar field potential depends 
heavily on the conformal frame adopted and, accordingly, so 
do the fact that the potential has a minimum  and the 
notion of scalar field mass.  One has to live with 
this feature and its consequences  if one wants to switch 
conformal frame. Therefore, care must be taken when 
specifying the ``mass'' or ``range'' of  a Brans-Dicke-like 
scalar field in scalar-tensor gravity.

\section{An application to metric $f(R)$ gravity}

As an application, we consider metric $f(R)$ gravity described 
by the action~(\ref{f(R)action}). This class of theories was 
important in building models of inflation in the early 
universe, and is now widely used to model the current 
acceleration of the universe as an alternative to a mysterious 
dark energy \cite{CCT, CDTT}. In the metric formalism, in which 
the metric is the only variable and the connection is the metric 
one, the field equations are
\be
f'(R)R_{ab}-\frac{f(R)}{2}\, g_{ab}=\nabla_a\nabla_b 
f'(R)-g_{ab} \Box  f'(R) +8\pi G T_{ab}^{(m)} 
\ee
where a prime denotes differentiation with respect to $R$. It is 
required that $f'(R)>0$ in order to have a positive effective 
gravitational coupling  $G_{eff}=G/f'(R)$ and 
$f''(R)>0$ for local stability \cite{mattmodgrav}. 

It is well known that, if $ f'' \neq 0$, $f(R)$ gravity is 
equivalent to a scalar-tensor theory~\footnote{This  
is true for 
both metric and Palatini modified gravity: however, Palatini 
$f(R)$ gravity has been shown to be non-viable on many grounds 
\cite{BarausseSotiriouMiller, TremblayFaraoni, 
Iglesiasetal, PalatiniPLB} and we do not discuss it here.}  
\cite{STequivalence}. In fact, 
by introducing an auxiliary field $\chi$, the action
\be\label{equivalentaction}
S=\frac{1}{16\pi G}\int d^4 x \, \sqrt{-g} \left[ 
f(\chi)+f'(\chi)\left( R-\chi \right) +{\cal L}^{(m)} \right] 
\ee
is dynamically equivalent to~(\ref{f(R)action}). This is trivial 
if $\chi=R$. Vice-versa, variation of~(\ref{equivalentaction}) 
with respect to $\chi$ yields $ f''(\chi)\left(R-\chi\right)=0$, 
and 
$\chi=R$ if $f''\neq 0$. We can now redefine the field $\chi$ as 
in $\phi \equiv f'(\chi)$ (this is now dimensionless). 
By setting
\be\label{effV}
V(\phi) = \chi(\phi) \phi -f(\chi(\phi)) \;,
\ee 
the action becomes
\be
S=\frac{1}{16\pi G} \int d^4x \, \sqrt{-g}\left[ \phi 
R-V(\phi)+{\cal L}^{(m)} \right] \;,
\ee
which describes an $\omega=0$ Brans-Dicke theory. The dynamical 
field $\phi=f'(R)$ obeys the trace equation
\be \label{traceeq}
3\Box \phi +2V(\phi)-\phi\, \frac{dV}{d\phi}=8\pi G T^{(m)} \;.
\ee
One is then led to establish the value of the mass of $\phi$ 
and to study the dynamics and stability of cosmological models 
based on this $f(R)$ gravity. Clearly, it would be incorrect to 
do this by looking at the shape, maxima, and minima of 
$V(\phi)$. The dynamics of $\phi$, through eq.~(\ref{traceeq}),  
are not regulated by $dV/d\phi$ but by the combination 
$\left[ \phi V'(\phi)-2V(\phi) \right]/3 $. Consider, for 
example, the model 
$f(R)=R+aR^2$ of interest for Starobinsky inflation in the early 
universe \cite{Starobinsky80, Starobinsky}. By naively taking the 
potential~(\ref{effV}), one obtains
\be
V(\phi)=a R^2 =\frac{1}{4a} \, \left( \phi -1 \right)^2 
\ee
and $\mu = 1/\sqrt{2a}$. The true mass $m$  of 
$\phi$ is instead given by
\be
m^2 = \frac{d^2V_{eff}}{d\phi^2} \;,
\ee
where 
\be \frac{dV_{eff}}{d\phi}=\frac{1}{3} \left[ \phi 
\frac{dV}{d\phi}-2V(\phi) \right] \;.
\ee
This leads to the correct mass $m=1/\sqrt{6a}$ (in agreement, 
{\em e.g.}, with \cite{Starobinsky80, Starobinsky91}). For a 
general 
$f(R)$ Lagrangian, it yields
\be
m^2 =\frac{d^2 V_{eff} }{d\phi^2}=\frac{f'-Rf''}{3f''} \;.
\ee
This expression has been derived in different contexts and with 
different methods: in various studies of stability 
\cite{FaraoniNadeau, Zerbini, Rador, 
NavarroVanAcoleyen}, perturbations  
\cite{SongHuSawicki, PogosianSilvestri}, 
propagator calculations for $f(R)$ 
gravity \cite{NunezSolganik}, {\em etc.} 
\cite{Odintsov0804, TsujikawaUddinTavakol, Olmo07, 
ChibaSmithErickcek}.

\section{Conclusions}

The notions of effective mass and range of a Brans-Dicke-like 
scalar 
field in scalar-tensor and $f(R)$ gravity are often used, 
especially in 
relation with the PPN formalism. However, the ambiguities 
related to the usual meaning of these concepts, induced  by 
familiarity with the Klein-Gordon equation instead of the 
wave equation obeyed by this field, are 
potentially dangerous. We have clarified this issue and pointed 
out the correct concept of scalar field effective mass.

A second issue is that of the representation-dependence of 
the scalar field potential that is used to define the  
effective mass. While, in classical physics, 
it is perfectly legitimate to switch conformal frames, care must 
be 
taken to ensure that the definitions and 
transformations of geometrical and physical quantities are 
implemented properly. Even without switching frame, 
attention has been called to certain potentials 
(fortunately only  a few of them) which can lead to 
errors. Unfortunately the terminology currently in use 
can, and does, lead to misunderstandings. 
While, at the classical  level, we have identified the Jordan frame 
mass $m$ as the 
appropriate definition,  when one tries to quantize gravity, one needs 
to resort to 
the Einstein frame quantities,  and the mass $\tilde{m}$ is then 
singled out. It is  not surprising that  one is faced with this change 
of prescription: it is well-known 
that, while equivalent at the classical level, these two
frames become inequivalent at the quantum level 
(\cite{FaraoniNadeauconftransf, STV} and references therein).

In addition to the definition of mass of the Brans-Dicke 
scalar,  the definition of cosmological constant 
deserves  a comment. A constant term $2\Lambda$ in the scalar 
field 
potential $V$ corresponds to a cosmological constant 
for non-minimally coupled scalars. However, due to the role 
played by $V$ in the Brans-Dicke field 
equation~(\ref{BDfieldeq2}),  this is no longer true in 
scalar-tensor gravity (again, we limit 
ourselves to Brans-Dicke theory for the sake of illustration). 
Setting $V=V_0=$const. does not make $V$ disappear from this 
equation. Moreover, if one wants to add a term $-\Lambda g_{ab}$ 
to the right hand side of the field equation~((\ref{BDfieldeq1})
 for the metric tensor, one should 
contemplate the linear potential $V=2\Lambda \phi$, since the  
relevant term in this equation is $-\, \frac{V}{2\phi} \, 
g_{ab}$. It is $V=2\Lambda \phi$, and not $V=2\Lambda$ that 
produces  a $-\Lambda g_{ab}$  term in~(\ref{BDfieldeq1}). The 
linear potential  produces an effective mass squared 
$m^2=\frac{-2\Lambda}{2\omega+3} $ for the field $\phi$, as 
discussed in Sec.~2.

\section*{Acknowledgment} 
This work is supported by the Natural  Sciences and Engineering Research Council of 
Canada (NSERC).

\end{document}